\documentclass{aastex61}

\shorttitle{ZTF Alert Distribution System}
\shortauthors{Patterson et al.}

\begin{document}

\title{The Zwicky Transient Facility Alert Distribution System}

\correspondingauthor{Maria T. Patterson}
\email{mtpatter@uw.edu}

\author[0000-0002-4753-3387]{Maria T. Patterson}
\affiliation{DIRAC Institute, Department of Astronomy, University of Washington, 3910 15th Avenue NE, Seattle, WA 98195, USA}

\author[0000-0001-8018-5348]{Eric C. Bellm}
\affiliation{DIRAC Institute, Department of Astronomy, University of Washington, 3910 15th Avenue NE, Seattle, WA 98195, USA}

\author[0000-0001-7648-4142]{Ben Rusholme}
\affiliation{IPAC, California Institute of Technology, 1200 E. California
             Blvd, Pasadena, CA 91125, USA}

\author[0000-0002-8532-9395]{Frank J. Masci}
\affiliation{IPAC, California Institute of Technology, 1200 E. California
             Blvd, Pasadena, CA 91125, USA}

\author[0000-0003-1996-9252]{Mario Juric}
\affiliation{DIRAC Institute, Department of Astronomy, University of Washington, 3910 15th Avenue NE, Seattle, WA 98195, USA}

\author[0000-0002-4410-7868]{K. Simon Krughoff}
\affil{LSST Project Office, 950 N. Cherry Avenue, Tucson, AZ 85719, USA}

\author[0000-0001-8205-2506]{V. Zach Golkhou}
\altaffiliation{Moore-Sloan, WRF, and DIRAC Fellow}
\affiliation{DIRAC Institute, Department of Astronomy, University of Washington, 3910 15th Avenue NE, Seattle, WA 98195, USA}
\affiliation{The eScience Institute, University of Washington, Seattle, WA 98195, USA}

\author[0000-0002-3168-0139]{Matthew J. Graham}
\affiliation{Division of Physics, Mathematics, and Astronomy, California Institute of Technology, Pasadena, CA 91125, USA}

\author[0000-0001-5390-8563]{Shrinivas R. Kulkarni}
\affiliation{Division of Physics, Mathematics, and Astronomy, California Institute of Technology, Pasadena, CA 91125, USA}

\author[0000-0003-3367-3415]{George Helou}
\affiliation{IPAC, California Institute of Technology, 1200 E. California
             Blvd, Pasadena, CA 91125, USA}

\collaboration{Zwicky Transient Facility Collaboration}



\begin{abstract}

The Zwicky Transient Facility (ZTF) survey generates real-time alerts for optical transients, variables, and moving objects discovered in its wide-field survey.
We describe the ZTF alert stream distribution
and processing (filtering) system.  
The system uses existing open-source technologies developed in industry: Kafka, a real-time streaming platform, and Avro, a binary serialization format.
The technologies used in this 
system provide a number of advantages for the ZTF use case, including 1) built-in 
replication, scalability, and stream “rewind” for the distribution mechanism, 2) 
structured messages with strictly enforced schemas and dynamic typing for fast 
parsing, and 3) a Python-based stream processing interface that is similar to 
batch for a familiar and user-friendly plug-in filter system, all in a modular, 
primarily containerized system. 
The production deployment has successfully supported streaming up to 1.2 million
alerts or roughly 70 GB of data per night, with each alert available to a 
consumer within about 10 seconds of alert candidate production. 
Data transfer rates of about 80,000 alerts/minute have been observed. 

In this paper, we discuss this alert distribution 
and processing system, the design motivations for the technology 
choices for the framework, performance in production, and how this system may be generally suitable for other 
alert stream use cases, including the upcoming Large Synoptic Survey Telescope 
(LSST).

\end{abstract}




\section{Introduction} \label{sec:intro}

The Zwicky Transient Facility \citep[ZTF;][]{tmp_Bellm:18:ZTFOverview,tmp_Graham:18:ZTFScience} distributes a live stream of its detections of 
transient astronomical events at a rate of 
about 1 million alerts per night. 
Contemporaneous observations 
by different types of telescopes and instruments at a 
range of wavelengths is essential to understand these astrophysical 
phenomena \citep[e.g.,][]{2014Natur.509..471G, 2015Natur.521..328C,2015ApJ...803L..24C}. 
In the past, a major limiting factor in securing followup
observations of transient events was the  
delay between the initial detection of an object by one 
telescope and reporting to other telescopes. 

The use of electronic distribution mechanisms such as the GRB Coordinates Network (GCN), Astronomer’s Telegrams \citep[ATEL;][]{Rutledge1998} , IAU Circulars, the Transient Name Server\footnote{\url{https://wis-tns.weizmann.ac.il/}},
and services relaying Virtual Observatory Event \citep[VOEvent;][]{Williams2006} messages have 
significantly alleviated the dissemination bottleneck. 
These mechanisms 
provide fast dissemination of transient alerts, but the data volume to date has been 
manageable enough for manual and visual processing upon receipt. 
For reference, 
the 4 Pi Sky collaboration maintains a database of VOEvent alert events going 
back to April 2014 that includes about 2.2 million alerts as of the writing of 
this paper (see \citealp{Staley2016} and \citealp{4pi}).
The largest volume of these alerts are reports of \textit{Swift} re-pointings broadcast by the GCN,
but they also include photometric alerts from the ASAS-SN and \textit{Gaia} collaborations among others.

However, in the era of ZTF and upcoming 
Large Synoptic Survey Telescope \citep[LSST;][]{Ivezic2008}, the bottleneck to discovery 
will likely shift from distribution to the ability to quickly isolate and 
prioritize detections of interest amid the flood of alert events:
The volume of VOEvents included in the 4 Pi Sky database from all sources is about two nights of alerts from ZTF.
Rather than only reporting events likely to be new explosive extragalactic transients, these surveys will stream \textit{all} sources that are above a specified detection threshold in the difference image, whether they are likely due to transients, variable stars, or moving objects.
This approach allows the science user the greatest flexibility in identifying candidates of interest for their individual science program.
However, these alert streams place greater demands on network bandwidth and require new software systems to filter and aggregate sources of interest.
Robust and 
scalable real-time processing capabilities for automated filtering and analysis of 
these alerts and subsequent integration capabilities with distribution mechanisms 
is therefore critical for enabling fast follow-up action and observations of time sensitive 
astronomical activity.

For ZTF project pipeline needs, we describe here the technologies used by the
ZTF Alert Distribution System (ZADS) for 
streaming these transient alert data, the motivations around the choices for the 
framework, the performance of the system, and how this system may be scaled for 
larger astronomical data stream needs.

\section{ZADS Overview and Goals}
\begin{figure}[!ht] 
\centering
\includegraphics[width=0.80\textwidth]{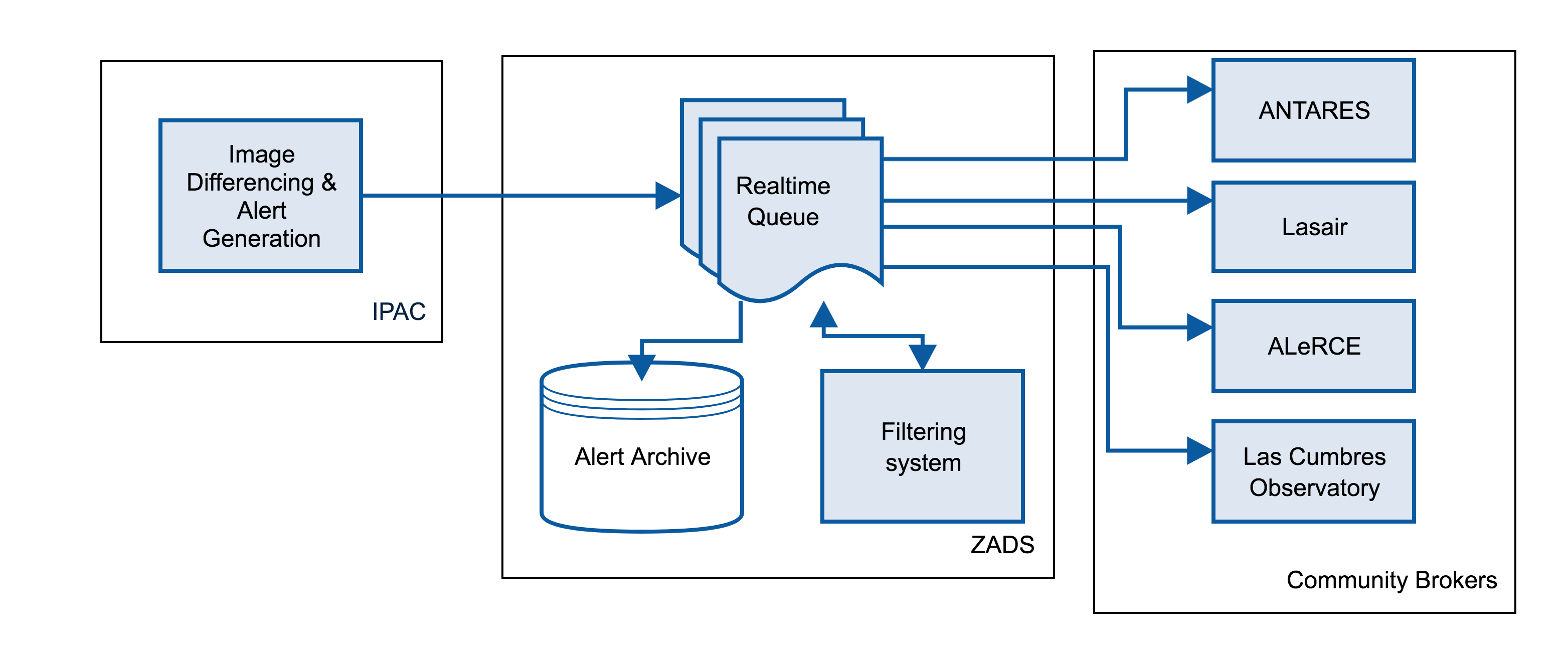}
\caption{Schematic overview of ZTF alert flow.
\label{fig:overview}}
\end{figure}

The ZTF Alert Distribution System is the near-real-time streaming platform that 
allows fast access to and filtering of the data products from ZTF's 
moving and varying object detection pipeline.
ZADS is a component of the ZTF Science Data System
\citep{tmp_Masci:18:ZTFDataSystem} that obtains alerts from  
the image differencing object detection and alert generation process and delivers them
to downstream brokers and science users (Figure \ref{fig:overview}).
The goal of ZADS is to support the availability of science quality alerts from ZTF
within 20 minutes of observation.

Given the high detection alert rate of ZTF, ZADS must be robust enough to support 
the streaming of a relatively large amount of data - over 1 million alerts per night, 
each with message size of around 60 KB for a total volume over 70 GB of nightly data.
The alerts must be moved out of IPAC's system and support delivery to the alert archive 
housed at University of Washington, to a public facing cloud system, 
and to downstream community event brokers with no data loss.
The system currently feeds four main event broker systems which further distribute
and provide access to science users: the Arizona NOAO
Temporal Analysis and Response to Events System \citep[ANTARES;][]{2018ApJS..236....9N}; Lasair\footnote{\url{http://lasair.roe.ac.uk/}}, a UK-based broker;
an initiative based in Chile called Automatic Learning for the Rapid Classification of 
Events (ALeRCE)\footnote{\url{http://alerce.science/}}; and Las Cumbres Observatory's Make Alerts Really Simple (MARS) project\footnote{\url{https://mars.lco.global/}}.
Translation to a VOEvent stream over VTP or automating the distribution 
of Gamma-ray Burst Coordinates Network (GCN) circulars or ATELs has not been attempted by ZADS,
but automatic reporting of candidate supernovae to the Transient Name Server (TNS)
is planned.

The ZADS platform is key to enabling real-time astronomical time domain science, 
including the discovery of young supernovae, the detection of near-Earth asteroids, 
the identification of stellar variables, and the search
for electromagnetic counterparts to gravitational wave sources.

\section{Technologies for ZADS}

\subsection{Alert formatting}

A new bottleneck to discovery in the flood of alert events is the ability to
quickly prioritize interesting alerts, necessitating
real-time filtering capabilities for the automated detection of objects of 
interest.
The primary alert format used by a number of projects follows the 
current IVOA standard VOEvent 2.0 \citep{Seaman2011}, which is an XML-based,
semi-structured data format that is appropriate for both human interaction and machine-readable scenarios.
The XML format typically results in an alert package of larger size compared to other encoding schemes given its redundant tags,
and filtering of VOEvents requires significant parsing.
Given the needs of the system, the Zwicky Alert Distribution System uses a more 
structured data format that is smaller and more suitable for fast parsing and
filtering.

ZADS transmits alerts using Apache Avro\footnote{\url{http://avro.apache.org/}}, a binary format developed in the 
open source Hadoop ecosystem that uses JSON-based schemas in its serialization 
framework.
Using Avro allows for a smaller sized alert message, with on average a factor of 
six smaller size compared with XML, and also faster serialization and 
deserialization, with a speedup observed to be approximately 40 times
\citep{Maeda2012}.
ZTF Avro alerts are approximately half the size of the same alert content in XML 
format with schema embedded and one-third the size without schema.
When packaging many alerts in the same .avro file, only one schema is needed
for all alerts.
Avro alerts are two-thirds the size of compressed XML.

Avro client libraries are available for many major programming languages, 
including Python 2 and 3.
However, we have found that these libraries can vary in quality, at least for 
Python 3, particularly with a trade-off between speed and functionality.

The ZTF alert schemas are defined in JSON documents in which the alert contents
are described as fields with a name, a data type (e.g., string, int, long, array,
null), and an optional documentation string for human reading.  
The data types are strictly adhered to and are correctly interpreted when deserialized
by the data receiver.  The schemas can also define
and allow for serializing fields as type ``bytes," which allows for the 
inclusion of binary format data such as individual files.  This makes it possible
to embed cutout FITS images of the science, reference, and difference images of
detected sources in a much more compact way than can be done with the current 
VOEvent standard.
The schemas used by the writer (ZTF project) and the reader (downstream brokers)
can also be different, allowing for the easy addition or removal of fields upon 
receipt.

ZTF makes use of the ability to ``nest" schemas in building the alert format, 
which offers some advantages over the current data model of VOEvent.
The main contents of the alert packets fall under the namespace {\tt\string ztf.alert}
as shown in Table \ref{tab:schema}.
Below {\tt\string ztf.alert} are the primary components including fields for 
object identification ({\tt\string ztf.alert.objectId}), the alert candidate 
detection data ({\tt\string ztf.alert.candidate}), and image cutouts 
(each of type {\tt\string ztf.alert.cutout}).
The alert candidate detection history is of type array, ({a list of \tt\string ztf.alert.prv\_candidate} all with the same schema), which can be of arbitrary length 
and is easily extensible as the detection history grows.

Note that the current ZADS Avro message format and data model was designed to fulfill the 
needs of the ZTF survey specifically. For example, candidate detection history is 
formatted in such a way as to allow for the simple processing of lightcurves,
and ZTF alerts include postage stamp cutout files as opposed to the URI to a
location where they can be accessed.
By comparison, VOEvent is intended as a more generalized format and attempts 
to specify a data model suitable for any celestial event use case.
Fulfilling the needs of ZTF stream and merging its alert structure 
with the more broadly applicable VOEvent could provide a roadmap for evolving IVOA standards
in preparation for alert streams from large scale surveys such as the LSST.  

\subsection{Alert message distribution}

For distributing alerts, ZADS utilizes Apache Kafka, a logging system or messaging
queue reinvented as a distributed data streaming platform \citep{Kreps2011}.
Kafka was originally developed by LinkedIn and later open sourced through the 
Apache Software Foundation.  Kafka is used for large scale, high throughput and
low latency data pipelines, has been deployed for industry applications 
transmitting over 1 trillions messages per day, 
and can be scaled easily by deploying a cluster of Kafka ``brokers.''  ZADS 
deploys Kafka in a small cluster of three brokers, using Docker\footnote{\url{https://www.docker.com}}, a software 
containerization platform.

Messages are sent to Kafka message queues by ``producers'' and separated into
partitions within a ``topic,'' which can be thought of as a named single stream 
which can be subscribed to or a single log that can be sequentially read from.  
The alert messages are ordered in the partitions 
and given an offset and a timestamp.  Downstream ``consumers'' subscribe to 
individual topics.  Instead of keeping one copy of a stream per consumer,
Kafka instead tracks the latest message offset read from a topic, making it scale
easily with the addition of new consumers.
This offset tracking method provides the advantage that the offsets can be 
changed so that consumers can seek to previous alert messages and receive 
alerts sent with timestamps prior to connecting to the stream.
This ensures that those listening to the stream will not miss alert messages
and also allows consumers to rewind to past offsets in order to reprocess data.
This rewind feature is not available in e.g., services deploying the current
transfer system of VOEvents, the VO Event Transport Protocol 
\citep[VTP;][]{Allan2017}

Alert consumers may wish to copy an entire topic or topics from the ZADS Kafka
system to a locally deployed Kafka cluster from which downstream processing can be
done.  This consumer is essentially a downstream mirror of topics from the 
original cluster.  Kafka's ``MirrorMaker'' feature makes this process of 
replicating streams straightforward, by deploying a consumer of topics from the 
source cluster and a producer to the target cluster. 

In addition to MirrorMaker, there are a variety of tools in the Kafka ecosystem, 
which we do not as of yet make use of.  Kafka Connect allows users to simply define sources 
of streams and sinks to external databases and other storage systems.
Kafka Streams is an internal pipeline development library, but it does not currently
support Python.
KSQL enables data processing with a SQL-like language and supports data of
type JSON and Avro.  However, use with ZTF's Avro formatted data would also require 
Kafka's schema registry system, which we have not yet explored.

Though we have been able to successfully put Kafka into production, the deployment
has not been without some degree of difficulty.
As a relatively young technology, it is not commonplace, particularly for 
science use cases, and expertise in the community is somewhat lacking.
Confluent, the company founded by the team who first built Kafka, does, however, 
provide a Slack team for discussion currently with over 6,000 users.
We have also found that the documentation for Kafka, though extensive and revised
for each distribution, can occasionally be rather terse.
Authentication and authorization of users and access control is available but 
difficult to set up.
Additionally, accounting support within Kafka, which would allow, e.g., monitoring
of data transfer to clients, is unavailable as an open source product.

\subsection{Filtering}

ZADS is in the process of deploying pre-defined filters of the stream for 
a number of science use cases.
These pre-defined filters are to be deployed using Python code in Docker containers.
Each filter reads from a Kafka topic and produces to a separate Kafka topic.
Alerts are pulled into a filter and deserialized from Avro into Python 
dictionaries.  The dictionaries then pass through the if-then logic of a 
simple Python function that returns True if the alert is of interest
and False if not.
Alerts that pass the filter are then serialized again to Avro and pushed 
back to a Kafka topic, for downstream consumption.

The above process can also be employed for user-defined filters.
In order to build reasonable working filters, a user might find a sample
set of batch alerts useful.  A batch of alerts can be collected by 
consuming from the stream and writing each Avro packet out to files on disk.
The batched Avro packets can be processed with the same filtering code used 
to process alert Avro packets received in the live stream, given that the 
data structure is the same for both batch and stream processing.
Users can then develop filtering code using batch data as a training set and 
deploy filters on the live stream with few changes to the code.

\section{ZADS Deployment in ZTF Nightly Processing}

\begin{figure}[!ht] 
\centering
\includegraphics[width=0.80\textwidth]{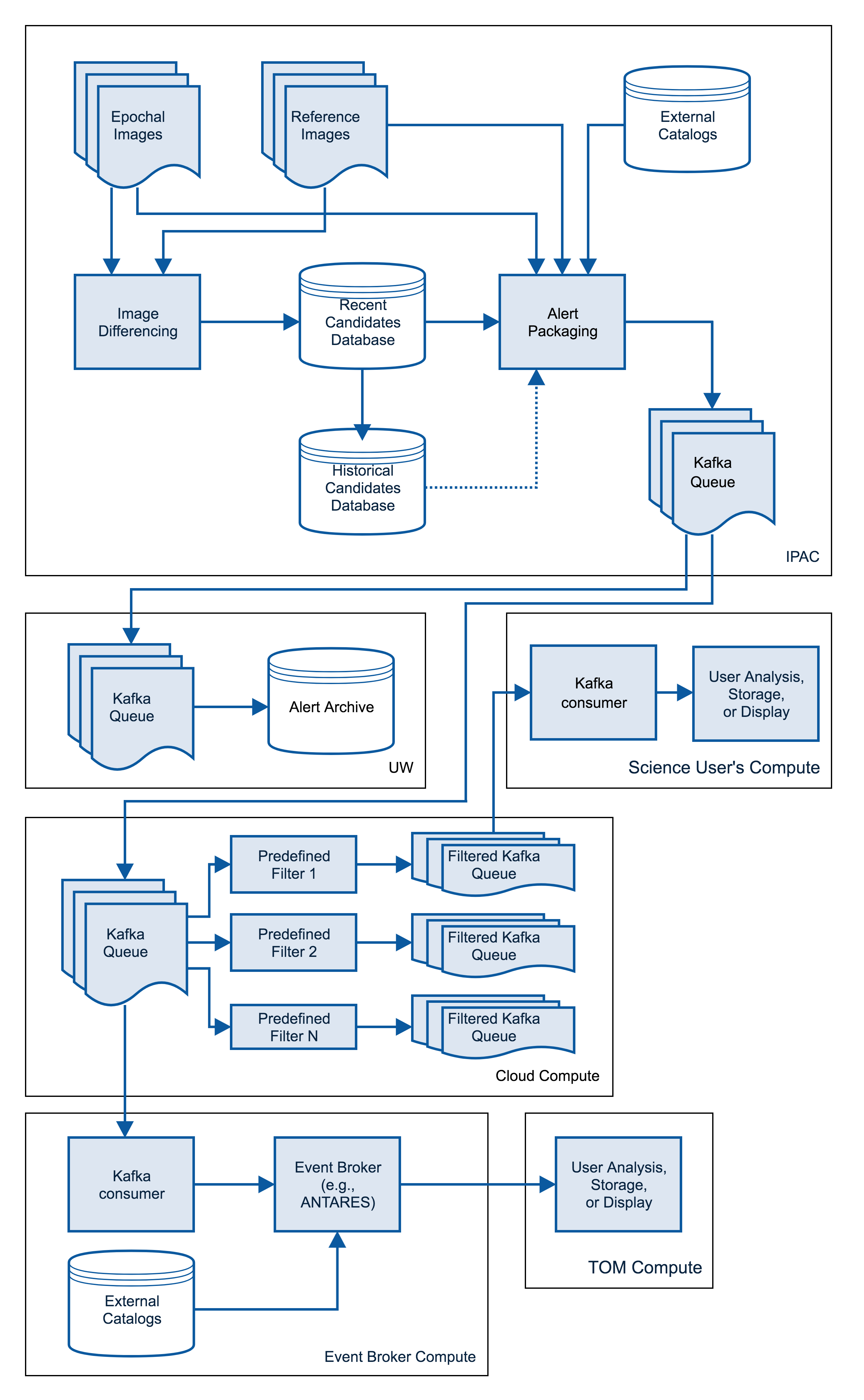}
\caption{ Diagram of the flow of alert data through the ZTF Alert Distribution
System.}
\label{pipeline}
\end{figure}

A diagram of the flow of ZTF alert data through the ZADS pipeline is shown 
in Fig.\ref{pipeline}.
The nightly alert pipeline begins at IPAC at Caltech when new detections 
are produced from image differencing between epochal and reference image data.
The candidates are loaded into a database where they are cross-matched 
to a database of previous historical ZTF candidates and to the Pan-STARRS1 and \textit{Gaia}
catalogs.  
Alerts are packaged into Avro format, containing, e.g., brightness, 
time, position, image cutouts of before, event, and subtraction, 
cross-matching to other catalogs, and previous detection information.
The full schema as well as sample alert packets are available\footnote{\url{https://github.com/ZwickyTransientFacility/ztf-avro-alert}}.
Table \ref{tab:schema} provides a high-level summary of some of the fields included
in the alert packets.
The pipeline produces Avro alerts in parallel, using Kafka as a buffer, reducing
to fewer streams, or Kafka topics, divided by date
and by program ID. This separates internal project collaboration 
alerts and public alerts into separate streams, which are then ready for further 
distribution outside of IPAC.

Alerts from moving objects are included in the stream, but Minor Planet Center 
reporting is handled separately from ZADS, as ZADS is a general purpose
alert distribution mechanism and interfaces with science-specific interfaces 
require additional filtering and post-processing.
Candidate moving objects are submitted to the MPC after a Moving Object 
Processing System (MOPS) process runs.
More detail on moving object handling is provided in the ZTF Data System 
overview paper \citep{tmp_Masci:18:ZTFDataSystem}.

\begin{deluxetable*}{lll}[h!]
\tablecaption{Partial schema of ZTF Avro alert packets\label{tab:schema}}
\tablecolumns{3}
\tablenum{1}
\tablewidth{0pt}
\tablehead{
\colhead{Field} & 
\colhead{Type} &
\colhead{Contents} 
}
\startdata
objectId & long & unique identifier for this object  \\
candid & long & unique identifier for the subtraction candidate \\
candidate & ztf.alert.candidate & nested schema record including fields prefixed by ``candidate" below \\
candidate.fid & int & filter ID (1=g; 2=r; 3=i) \\
candidate.ra  &	double & Right Ascension of candidate; J2000 [deg] \\
candidate.dec & double & Declination of candidate; J2000 [deg] \\
candidate.magpsf & float & magnitude from PSF-fit photometry [mag] \\
candidate.distnr & float or null & distance to nearest source in reference image PSF-catalog within 30 arcsec [pixels] \\
candidate.magnr	& float or null & magnitude of nearest source in reference image PSF-catalog within 30 arcsec [mag] \\
candidate.classtar & float or null  & star/galaxy classification score of candidate from SExtractor \\
candidate.rb & float or null & RealBogus quality score; range is 0 to 1 where closer to 1 is more reliable \\
prv\_candidate & array of ztf.alert.candidate & associated alert candidate records for the last 30 days of history \\
cutoutScience & ztf.alert.cutout or null & cutout of the science image \\
cutoutTemplate & ztf.alert.cutout or null & cutout of the co-added reference image \\
cutoutDifference & ztf.alert.cutout or null & cutout of the resulting difference image \\
\enddata
\tablecomments{ZTF uses nested schemas to organize the data in the alert packet.
{\tt\string ztf.alert} is the top-level namespace, the contents of which are shown above. {\tt\string ztf.alert} relies on 
nested schemas {\tt\string ztf.alert.candidate}, {\tt\string ztf.alert.prv\_candidate},
and {\tt\string ztf.alert.cutout}.
The {\tt\string prv\_candidate} field contains an array of one or more previous subtraction candidates at the position of the alert. These are obtained by a simple cone search at the position of the alert candidate on the last 30 days of history. If there are no previous candidates or upper limits, this field is null.
The fields for an individual {\tt\string prv\_candidate} are nearly identical to candidate except for the omission of the PS1 matches, previous detection history, and reference image information.
}
\end{deluxetable*}

From IPAC, alerts are pulled into another instance of Kafka at the University of
Washington (UW) via Kafka's MirrorMaker functionality.  The UW Kafka instance 
runs as a small cluster of three Kafka brokers in Docker containers, mirroring 
collaboration and public program alert stream topics.  From the UW Kafka instance,
a local consumer reads each night of data and saves the stream to an alert 
archive.  This archive is primarily for backup purposes and to provide limited 
access to past alerts in case of connection issues with community brokers.
While more capable community broker tools are in development, we are also using this archive to provide community access to public alerts as bulk downloads\footnote{\url{https://ztf.uw.edu/alerts/public/}}.
The UW Kafka instance also distributes alerts to ZTF internal collaboration
partners.

The IPAC cluster also feeds the public stream into a cloud-based Kafka instance, 
populated again by using the MirrorMaker functionality. This Kafka instance is
currently running on the DigitalOcean cloud infrastructure. 
The public stream is then available to astronomical event ``community brokers",
Target and Observation Managers (TOMs), and similar systems through the cloud-based 
Kafka instance.  
Both the cloud-based Kafka broker and the UW Kafka broker keep a 7-day ``hot cache''
of alert streams, making one week of past alerts available to community brokers 
in addition to the live stream.
Alert stream topics are separated into 16 partitions.  
This allows consumers of the stream to 
parallelize readers into a group of up to 16, allowing faster reading 
and parallel downstream filtering and processing.

The cloud-based Kafka system will also feed into pre-defined community filters.
These pre-defined community filters will run co-located in the cloud.
Community filters are intended to be a first pass at detecting and classifying,
for example, supernova or variable stars.
Alerts that pass a community filter are routed back to the Kafka instance
in a separate topic per filter. 
These filtered streams are then available for analysis by science users.

\section{Performance and metrics}

ZTF produces around 600,000 to 1.2 million alerts on a full night of observing.
The ZADS system has been observed to withstand a throughput of over 2 million
alerts with no technical issues.
The size of each individual alert is about 60 KB, dominated by the
size of the included cutout images.
The total volume of ZTF alerts pushed through ZADS over one night can then 
amount to over 70 GB of data.

The total end-to-end ZADS pipeline results in alerts available roughly 20 minutes
after a field is observed, dominated by the data reduction time.
The median time between alert candidate production and packaging to availability 
in the IPAC Kafka instance is about 6 seconds.
In simulations, the serialization of 1,000 ZTF alerts into Avro format and 
submission to a Kafka topic, and transfer to the consumer was observed to
be 4.2 seconds.
Data transfer rates of about 100 MBps to the cloud have been observed.
This equates to over 80k alerts/minute or 6.6k alerts in 5 seconds.

\section{Interface}

ZADS provides two main endpoints for scientific users to connect and receive
alerts.  One way to access the public stream is via
astronomical event ``community brokers."  Community brokers consume the full
public stream and provide value added services such as cross-matching 
with external catalogs to enrich alerts.
After community broker processing and additional filtering, scientific
users can access these enriched alerts via the broker interfaces.

Science users may also wish to directly access the public stream from 
their own compute resources.  At present, access to the public stream is expected to be limited to 
filtered topics from the pre-defined community filters.  Example code
demonstrating how an approved user can connect directly to a pre-defined community
filtered stream as a Kafka topic is available here:
\url{https://github.com/ZwickyTransientFacility/alert_stream}.
This code could also be used with any ZTF alert distribution hub that utilizes
a Kafka instance to stream alerts.
Note that this code is only an example and will not allow unknown users to connect 
to ZADS.
Connecting directly to ZADS is only available to broker projects who have signed
MOU agreements.
Access to the stream is currently restricted by IP address.
Encryption, authentication, and additional restrictions on consumers (topic 
access controls and throttling stream consumption) are available in Kafka
and planned to be implemented.
These features will be essential in future projects particularly with many consumers.

\section{Summary} 

We have described here the ZTF Alert Distribution System and the 
technologies used to implement the pipeline.
ZTF Avro formatted alerts are streamed from IPAC to the University of 
Washington as well as a cloud computing service and subsequently to downstream 
community brokers and scientific users via the Apache Kafka data streaming
platform.
Along the pipeline, alerts are collected to an alert archive and 
also processed with simple pre-defined community filters as a first pass 
at creating filtered streams separated into various distinct astronomical 
sources.

The ZADS platform has successfully transmitted more alerts than have ever
been distributed via VOEvents, on the order of one million alerts
nightly.
This is the first successful demonstration of at-scale alert 
distribution in an astronomical context, using industry standard tools and 
serialization formats, and may provide guidance for the evolution of 
astronomical standards in preparation for the Large Synoptic Survey Telescope.
The alert volume of ZTF approaches $~10\%$ of LSST, with significantly more modest
hardware used for ZADS; we therefore feel this is strong evidence to the LSST 
alert problem being tractable.
The robustness of the distribution system and filtering capabilities 
make ZADS promising as a precursor to the high volume of alerts expected from
the upcoming LSST. 




\acknowledgements
\section*{Acknowledgments} 

Based on observations obtained with the Samuel Oschin Telescope 48-inch and the 60-inch Telescope at the Palomar Observatory as part of the Zwicky Transient Facility project, a scientific collaboration among the California Institute of Technology, the Oskar Klein Centre, the Weizmann Institute of Science, the University of Maryland, the University of Washington, Deutsches Elektronen-Synchrotron, the University of Wisconsin-Milwaukee, and the TANGO Program of the University System of Taiwan. Further support is provided by the U.S.\ National Science Foundation under Grant No.\ AST-1440341.

M.~Patterson, E.~Bellm., and K.~S.~Krughoff acknowledge support from the Large Synoptic Survey Telescope program, which is supported in part by the National Science Foundation through
Cooperative Agreement 1258333 managed by the Association of Universities for Research in Astronomy
(AURA), and the Department of Energy under Contract No. DE-AC02-76SF00515 with the SLAC National
Accelerator Laboratory. Additional LSST funding comes from private donations, grants to universities,
and in-kind support from LSSTC Institutional Members.

M.~Patterson, E.~Bellm, V.~Z.~Golkhou, and M.~Juric
acknowledge support from the University of Washington College of Arts and Sciences, Department of Astronomy, and the DIRAC Institute. University of Washington's DIRAC Institute is supported through generous gifts from the Charles and Lisa Simonyi Fund for Arts and Sciences, and the Washington Research Foundation. 

E.~Bellm is supported in part by the NSF AAG grant 1812779 and grant \#2018-0908 from the Heising-Simons Foundation.

M.~Juric acknowledges the support of the Washington Research Foundation Data Science Term Chair fund, and the UW Provost's Initiative in Data-Intensive Discovery.

\bibliography{biblio}



\end{document}